\documentclass[apjl]{emulateapj}
\usepackage{comment}
\usepackage{ifthen}
\usepackage{lineno}

\newcommand{\forloop}[5][1]%
{%
\setcounter{#2}{#3}%
\ifthenelse{#4}%
	{%
	#5%
	\addtocounter{#2}{#1}%
	\forloop[#1]{#2}{\value{#2}}{#4}{#5}%
	}%
	{%
	}%
}%

\newcommand{\ctbd}[1]{}

\newcommand{\lc}{light curve}
\newcommand{\lcs}{light curves}
\newcommand{\Lc}{Light curve}

\newcommand{\band}[1]{\ensuremath{#1}~band}

\newcommand{\kms}{\ensuremath{\rm km\,s^{-1}}}
\newcommand{\ms}{\ensuremath{\rm m\,s^{-1}}}

\newcommand{\gcmc}{\ensuremath{\rm g\,cm^{-3}}}
\newcommand{\ergscmsq}{\ensuremath{\rm erg\,s^{-1}\,cm^{-2}}}

\newcommand{\vsini}{\ensuremath{v \sin{i}}}
\newcommand{\feh}{\ensuremath{\rm [Fe/H]}}

\newcommand{\rhk}{\ensuremath{R^{\prime}_{\rm HK}}}
\newcommand{\logrhk}{\ensuremath{\log\rhk}}
\newcommand{\Savg}{\ensuremath{\langle S\rangle}}

\newcommand{\rsun}{\ensuremath{R_\sun}}
\newcommand{\msun}{\ensuremath{M_\sun}}
\newcommand{\lsun}{\ensuremath{L_\sun}}

\newcommand{\rstar}{\ensuremath{R_\star}}
\newcommand{\mstar}{\ensuremath{M_\star}}
\newcommand{\lstar}{\ensuremath{L_\star}}

\newcommand{\teffstar}{\ensuremath{T_{\rm eff\star}}}
\newcommand{\rhostar}{\ensuremath{\rho_\star}}
\newcommand{\loggstar}{\ensuremath{\log{g_{\star}}}}

\newcommand{\rearth}{\ensuremath{R_\earth}}
\newcommand{\mearth}{\ensuremath{M_\earth}}

\newcommand{\rpl}{\ensuremath{R_{p}}}
\newcommand{\mpl}{\ensuremath{M_{p}}}

\newcommand{\rhopl}{\ensuremath{\rho_{p}}}

\newcommand{\gpl}{\ensuremath{g_{p}}}

\newcommand{\arstar}{\ensuremath{a/\rstar}}
\newcommand{\zrstar}{\ensuremath{\zeta/\rstar}}

\newcommand{\rjup}{\ensuremath{R_{\rm J}}}
\newcommand{\mjup}{\ensuremath{M_{\rm J}}}

\newcommand{\reffig}[1]{Fig.~\ref{fig:#1}}
\newcommand{\refsec}[1]{\mbox{\S\ \ref{sec:#1}}}

\newcommand{\reftab}[1]{Tab.~\ref{tab:#1}}

\newcommand{\refsecl}[1]{\mbox{Section \ref{sec:#1}}}

\newcommand{\reftabl}[1]{Table~\ref{tab:#1}}

\newcommand{\hd}[1]{\mbox{HD #1}}

\newcommand{\hatcurhtr}{HATS568-008}                                   
\newcommand{\hatcurCCra}{\ensuremath{13^{\mathrm h}55^{\mathrm m}25.68{\mathrm s}}}                                  
\newcommand{\hatcurCCdec}{\ensuremath{-21{\arcdeg}12{\arcmin}27.7{\arcsec}}}                                 
\newcommand{\hatcurCCtwomass}{13552567-2112276}                  
\newcommand{\hatcurCCgsc}{GSC~6148-00422}                              
\newcommand{\hatcurCCtassmv}{\ensuremath{13.340\pm0.010}}              
\newcommand{\hatcurCCtassmB}{\ensuremath{14.359\pm0.020}}              
\newcommand{\hatcurCCtassmg}{\ensuremath{13.825\pm0.010}}              
\newcommand{\hatcurCCtassmr}{\ensuremath{13.013\pm0.030}}              
\newcommand{\hatcurCCtassmi}{\ensuremath{12.690\pm0.030}}              
\newcommand{\hatcurCCtwomassJmag}{\ensuremath{11.528\pm0.024}}         
\newcommand{\hatcurCCtwomassHmag}{\ensuremath{11.085\pm0.024}}         
\newcommand{\hatcurCCtwomassKmag}{\ensuremath{10.976\pm0.026}}         
\newcommand{\hatcurLCrprstar}{\ensuremath{0.0711\pm0.0019}}            
\newcommand{\hatcurLCbsq}{\ensuremath{0.148_{-0.085}^{+0.101}}}        
\newcommand{\hatcurLCimp}{\ensuremath{0.39_{-0.13}^{+0.11}}}           
\newcommand{\hatcurLCzeta}{\ensuremath{22.62\pm0.16}}                  
\newcommand{\hatcurLCdur}{\ensuremath{0.0958\pm0.0012}}                
\newcommand{\hatcurLCingdur}{\ensuremath{0.00735\pm0.00099}}           
\newcommand{\hatcurLCP}{\ensuremath{3.1853150\pm0.0000054}}            
\newcommand{\hatcurLCPprec}{\ensuremath{3.1853150}}                    
\newcommand{\hatcurLCPshort}{\ensuremath{3.1853}}                      
\newcommand{\hatcurLCT}{\ensuremath{2456528.29697\pm0.00058}}          
\newcommand{\hatcurSMEiteff}{\ensuremath{5030\pm50}}                   
\newcommand{\hatcurSMEizfeh}{\ensuremath{+0.270\pm0.080}}               
\newcommand{\hatcurSMEizfehshort}{\ensuremath{+0.27}}                   
\newcommand{\hatcurSMEilogg}{\ensuremath{4.62\pm0.10}}                 
\newcommand{\hatcurSMEivsin}{\ensuremath{0.40\pm0.50}}                 
\newcommand{\hatcurSMEivmac}{\ensuremath{0.0}}                         
\newcommand{\hatcurSMEivmic}{\ensuremath{0.0}}                         
\newcommand{\hatcurSMEiiteff}{\ensuremath{4985\pm50}}                  
\newcommand{\hatcurSMEiizfeh}{\ensuremath{+0.250\pm0.080}}              
\newcommand{\hatcurSMEiizfehshort}{\ensuremath{+0.25}}                  
\newcommand{\hatcurSMEiilogg}{\ensuremath{4.530\pm0.050}}              
\newcommand{\hatcurSMEiivsin}{\ensuremath{0.50\pm0.50}}                
\newcommand{\hatcurSMElogrhk}{\ensuremath{-4.800}}                     
\newcommand{\hatcurSMES}{\ensuremath{0.27}}                            
\newcommand{\hatcurLBii}{\ensuremath{0.4140}}                          
\newcommand{\hatcurLBiii}{\ensuremath{0.2464}}                         
\newcommand{\hatcurLBir}{\ensuremath{0.5519}}                          
\newcommand{\hatcurLBiir}{\ensuremath{0.1984}}                         
\newcommand{\hatcurISOm}{\ensuremath{0.849\pm0.027}}                   
\newcommand{\hatcurISOmlong}{\ensuremath{0.849\pm0.027}}               
\newcommand{\hatcurISOr}{\ensuremath{0.815_{-0.035}^{+0.049}}}         
\newcommand{\hatcurISOrlong}{\ensuremath{0.815_{-0.035}^{+0.049}}}     
\newcommand{\hatcurISOrho}{\ensuremath{2.22\pm0.33}}                   
\newcommand{\hatcurISOlogg}{\ensuremath{4.545\pm0.049}}                
\newcommand{\hatcurISOlum}{\ensuremath{0.368_{-0.038}^{+0.050}}}       
\newcommand{\hatcurISOmv}{\ensuremath{6.10\pm0.14}}                    
\newcommand{\hatcurISOage}{\ensuremath{7.8\pm5.0}}                     
\newcommand{\hatcurISOMK}{\ensuremath{3.94\pm0.11}}                    
\newcommand{\hatcurISOspec}{K2}                                         
\newcommand{\hatcurRVK}{\ensuremath{18.4\pm1.9}}                       
\newcommand{\hatcurRVjitter}{\ensuremath{3.0\pm1.1}}                   
\newcommand{\hatcurPPi}{\ensuremath{87.92\pm0.75}}                     
\newcommand{\hatcurPPlogg}{\ensuremath{2.968\pm0.076}}                 
\newcommand{\hatcurPPar}{\ensuremath{10.59_{-0.67}^{+0.51}}}           
\newcommand{\hatcurPParel}{\ensuremath{0.04012\pm0.00043}}             
\newcommand{\hatcurPPrho}{\ensuremath{0.83\pm0.18}}                    
\newcommand{\hatcurPPm}{\ensuremath{0.120\pm0.012}}                    
\newcommand{\hatcurPPmlong}{\ensuremath{0.120\pm0.012}}                
\newcommand{\hatcurPPme}{\ensuremath{38.0\pm3.9}}                      
\newcommand{\hatcurPPr}{\ensuremath{0.563_{-0.034}^{+0.046}}}          
\newcommand{\hatcurPPrshort}{\ensuremath{0.56}}                        
\newcommand{\hatcurPPrlong}{\ensuremath{0.563_{-0.034}^{+0.046}}}      
\newcommand{\hatcurPPre}{\ensuremath{6.31_{-0.38}^{+0.52}}}            
\newcommand{\hatcurPPmrcorr}{\ensuremath{-0.02}}                       
\newcommand{\hatcurPPteff}{\ensuremath{1084\pm32}}                     
\newcommand{\hatcurPPtheta}{\ensuremath{0.0200\pm0.0024}}              
\newcommand{\hatcurPPfluxavg}{\ensuremath{3.12_{-0.32}^{+0.45}}}       
\newcommand{\hatcurXdist}{\ensuremath{261_{-12}^{+16}}}                
\newcommand{\hatcurXAv}{\ensuremath{0.186\pm0.058}}                    
\newcommand{\hatcurXdistred}{\ensuremath{257_{-12}^{+15}}}             
\newcommand{\hatcurCCpmra}{\ensuremath{-43.3\pm1.4}}                   
\newcommand{\hatcurCCpmdec}{\ensuremath{-37.2\pm1.9}}                  
\newcommand{\hatcurRVecceneccen}{\ensuremath{0.065\pm0.053}}           
\newcommand{\hatcurRVeccentwosiglimeccen}{\ensuremath{<0.170}}         
\newcommand{\hatcur}{HATS-7}
\newcommand{\hatcurb}{HATS-7b}
\newcommand{\hatcurRVgammaabs}{\ensuremath{9.435\pm0.017}}                           

\newcommand{\hatcurlumind}{\rhostar}
\newcommand{\hatcurjhkfilset}{ESO}

\newcommand{\hatcurSMEversion}{ii}                                       
\newcommand{\hatcurSMEteff}{\ifthenelse{\equal{\hatcurSMEversion}{i}}{\hatcurSMEiteff}{\hatcurSMEiiteff}}
\newcommand{\hatcurSMEzfeh}{\ifthenelse{\equal{\hatcurSMEversion}{i}}{\hatcurSMEizfeh}{\hatcurSMEiizfeh}}
\newcommand{\hatcurSMEzfehshort}{\ifthenelse{\equal{\hatcurSMEversion}{i}}{\hatcurSMEizfehshort}{\hatcurSMEiizfehshort}}
\newcommand{\hatcurSMElogg}{\ifthenelse{\equal{\hatcurSMEversion}{i}}{\hatcurSMEilogg}{\hatcurSMEiilogg}}
\newcommand{\hatcurSMEvsin}{\ifthenelse{\equal{\hatcurSMEversion}{i}}{\hatcurSMEivsin}{\hatcurSMEiivsin}}
\newcommand{\hatcurSMEvmac}{\ifthenelse{\equal{\hatcurSMEversion}{i}}{\hatcurSMEivmac}{\hatcurSMEiivmac}}
\newcommand{\hatcurSMEvmic}{\ifthenelse{\equal{\hatcurSMEversion}{i}}{\hatcurSMEivmic}{\hatcurSMEiivmic}}

\newboolean{emulateapj}
\setboolean{emulateapj}{true}

\newboolean{rvtablelong}
\setboolean{rvtablelong}{true}

\newboolean{astroph}
\setboolean{astroph}{true}

\shortauthors{Bakos et al.}
\shorttitle{
\hatcur\lowercase{b}
}
\ifthenelse{\boolean{emulateapj}}{
    \newcommand{\titledag}{$\dagger$}
}{
    \newcommand{\titledag}{\dagger}
}

\begin{document}
\title{
\hatcur\lowercase{b}: A Hot Super Neptune Transiting a Quiet K Dwarf Star 
\altaffilmark{\titledag}
}

\author{
	G.~\'A.~Bakos\altaffilmark{1,$\star$,$\star\star$},
	K.~Penev\altaffilmark{1},
    D.~Bayliss\altaffilmark{2,3}, 
    J.~D.~Hartman\altaffilmark{1},
    G.~Zhou\altaffilmark{2,1}, 
    R.~Brahm\altaffilmark{4,5},
    L.~Mancini\altaffilmark{6},   
    M.~de~Val-Borro\altaffilmark{1},
    W.~Bhatti\altaffilmark{1},
    A.~Jord\'an\altaffilmark{4,5},
	M.~Rabus\altaffilmark{4,6},
	N.~Espinoza\altaffilmark{4,5},
    Z.~Csubry\altaffilmark{1},    
    A.~W.~Howard\altaffilmark{7}, 
    B.~J.~Fulton\altaffilmark{7,8},
    L.~A.~Buchhave\altaffilmark{9,10},
    S.~Ciceri\altaffilmark{6}, 
    T.~Henning\altaffilmark{6},
    B.~Schmidt\altaffilmark{2},
    H.~Isaacson\altaffilmark{11},
    R.~W.~Noyes\altaffilmark{9},
    G.~W.~Marcy\altaffilmark{11},
    V.~Suc\altaffilmark{4},
	A.~R.~Howe\altaffilmark{1},
	A.~S.~Burrows\altaffilmark{1},
    J.~L\'az\'ar\altaffilmark{12},
    I.~Papp\altaffilmark{12},
    P.~S\'ari\altaffilmark{12}
}
\altaffiltext{1}{Department of Astrophysical Sciences, Princeton
  University, NJ 08544, USA} \altaffiltext{$\star$}{Alfred P.~Sloan
  Research Fellow} \altaffiltext{$\star\star$}{Packard Fellow}
\altaffiltext{2}{Research School of Astronomy and Astrophysics,
  Australian National University, Canberra, ACT 2611, Australia}
\altaffiltext{3}{Observatoire
  Astronomique de l'Universit\'e de Gen\`eve, 51 ch.~des Maillettes,
  1290 Versoix, Switzerland}
\altaffiltext{4}{Instituto de Astrof\'isica, Facultad de F\'isica,
  Pontificia Universidad Cat\'olica de Chile, Av. Vicu\~na Mackenna
  4860, 7820436 Macul, Santiago, Chile; rbrahm@astro.puc.cl}
\altaffiltext{5}{Millennium Institute of Astrophysics, Av. Vicu\~na
  Mackenna 4860, 7820436 Macul, Santiago, Chile}
\altaffiltext{6}{Max
  Planck Institute for Astronomy, Heidelberg, Germany}
\altaffiltext{7}{Institute for Astronomy, University of Hawaii at
  Manoa, Honolulu, HI, USA}
\altaffiltext{8}{NSF Graduate Research Fellow}
\altaffiltext{9}{Harvard-Smithsonian Center for Astrophysics,
  Cambridge, MA 02138, USA}
\altaffiltext{10}{Centre for Star and Planet Formation, Natural History
Museum, 
University of Copenhagen, 
Denmark}
\altaffiltext{11}{Department of Astronomy,
  University of California, Berkeley, CA 94720-3411, USA}
\altaffiltext{12}{Hungarian Astronomical
   Association, Budapest, Hungary}
\altaffiltext{$\dagger$}{
The HATSouth network is operated by a collaboration consisting of
Princeton University (PU), the Max Planck Institute f\"ur Astronomie
(MPIA), the Australian National University (ANU), and the Pontificia
Universidad Cat\'olica de Chile (PUC).  The station at Las Campanas
Observatory (LCO) of the Carnegie Institute is operated by PU in
conjunction with PUC, the station at the High Energy Spectroscopic
Survey (H.E.S.S.) site is operated in conjunction with MPIA, and the
station at Siding Spring Observatory (SSO) is operated jointly with
ANU.
This paper includes data gathered with the 10\,m Keck-I telescope at
Mauna Kea, 
the MPG~2.2\,m and ESO~3.6\,m
telescopes at the ESO Observatory in La Silla.  This
paper uses observations obtained with facilities of the Las Cumbres
Observatory Global Telescope.
}


\begin{abstract}

\setcounter{footnote}{10}
We report the discovery by the HATSouth network of \hatcurb{}, a
transiting Super-Neptune with a mass of \hatcurPPm\,\mjup, a radius of
\hatcurPPr\,\rjup, and an orbital period of \hatcurLCPshort\,days.
The
host star
%
%
is a moderately bright ($V=\hatcurCCtassmv$\,mag,
$K_{S}=\hatcurCCtwomassKmag$\,mag) K dwarf star with a mass of
\hatcurISOm\,\msun, a radius of \hatcurISOr\,\rsun, and a metallicity
of \feh$=\hatcurSMEzfeh$.  The star is photometrically quiet to within
the precision of the HATSouth measurements and has low RV jitter.
\hatcurb{} is the second smallest radius planet discovered by a
wide-field ground-based transit survey, and one of only a handful of
Neptune-size planets with mass and radius determined to 10\% precision. 
Theoretical modeling of \hatcurb\ yields a hydrogen-helium fraction of
$18\pm4$\% (rock-iron core and H$_2$-He envelope), or $9\pm4$\% (ice
core and H$_2$-He envelope), i.e.~it has a composition broadly similar
to that of Uranus and Neptune, and very different from that of Saturn,
which has 75\% of its mass in H$_2$-He.  Based on a sample of
transiting exoplanets with accurately ($<$20\%) determined parameters,
we establish approximate power-law relations for the envelopes of the
mass--density distribution of exoplanets.  \hatcurb{}, which, together
with the recently discovered HATS-8b, is one of the first two
transiting super-Neptunes discovered in the Southern sky, is a prime target
for additional follow-up observations with Southern hemisphere
facilities to characterize the atmospheres of Super-Neptunes (which we
define as objects with mass greater than that of Neptune, and smaller than
halfway between that of Neptune and Saturn, 
i.e.~$0.054\,\mjup < \mpl < 0.18\,\mjup$).
\setcounter{footnote}{0}
\end{abstract}



\section{Introduction}
\label{sec:introduction}

The HATSouth project \citep{bakos:2013:hatsouth} is the first global
network of homogeneous, fully automated telescopes, capable of
round-the-clock monitoring of a wide field-of-view (FOV) on the sky. 

The key scientific goal of HATSouth is to search for transiting
extrasolar planets (TEPs), especially ones that are smaller and/or have
longer periods than the hot Jupiters and Saturns, which have been the
primary type of planet found by ground-based surveys to date.  By
covering a wider total FOV than the NASA {\em Kepler}\/ mission
\citep{kepler}, HATSouth
monitors more bright stars per unit time than this mission, promising a
greater yield of planets amenable to detailed characterization. 

The HATSouth telescopes are installed at Las Campanas Observatory (LCO)
in Chile, at the H.E.S.S.~site in Namibia, and at Siding Spring
Observatory (SSO) in Australia.  HATSouth was commissioned in late
2009, with regular operations beginning in 2011 after a $\sim$1 year
shakedown period, and has since collected 2.9 million science images at
4\,minute cadence for some 9.5 million stars with $r < 16$\,mag (2\%
per-point precision) in the Southern sky.  As of May 1, 2015, the
HATSouth telescopes have opened on 1788, 1528, and 1324 nights from
LCO, HESS, and SSO, respectively.  Based on weather statistics through
2015, the sites have averaged 8.24\,hrs, 7.54\,hrs, and 5.24\,hrs of
useful dark hours per 24\,hr time period respectively.  The
longitudinal distribution of the three sites means that selected
regions of the sky may be observed for long contiguous stretches of
time, occasionally reaching 130 hours with interruptions shorter than
30 minutes.

Here we present the discovery of \hatcurb{}, which together with the
recently announced HATS-8b \citep{bayliss:15:hats8}, is one of the
first two transiting Super-Neptunes found by the HATSouth network, and
one of only four such planets discovered by a ground-based wide-field
survey (the other two being HAT-P-11b, \citealp{bakos:2010:hat11}, and
HAT-P-26b, \citealp{hartman:2011:hat26}).  By combining photometric
observations with high-resolution, high velocity-precision
spectroscopy, we determine the mass and radius of \hatcurb{} to better
than 10\% accuracy.  \hatcurb{} is one of only nine planets with $M <
0.18$\,\mjup\ for which both the mass and radius have been determined
to this level of accuracy\footnote{Based on the NASA exoplanet archive
\url{http://exoplanetarchive.ipac.caltech.edu} accessed 2015 March
25.}.  With $V = \hatcurCCtassmv$\,mag, \hatcur{} is the 7th brightest
star known to host a transiting planet with $M < 0.18$\,\mjup, and for
which the mass has been accurately (with 10\% accuracy) determined. 
And of these planets, only GJ~436 \citep{gillon:2007}, GJ~3470
\citep{bonfils:2012} and HAT-P-26 \citep{hartman:2011:hat26} have
deeper transits, making \hatcurb{} one of the best known low-mass
transiting planets for detailed follow-up studies, e.g.~transmission
spectroscopy for probing its atmospheric composition.

In the following section we describe the observations utilized to
discover and characterize \hatcurb{}.  In \refsec{analysis} we discuss
our analysis of the data to rule out blend scenarios and determine the
system parameters.  We discuss the results in \refsec{discussion}.

\section{Observations}
\label{sec:obs}

\subsection{Photometry}
\label{sec:photometry}

\subsubsection{Photometric detection}
\label{sec:detection}

\begin{figure}[]
\plotone{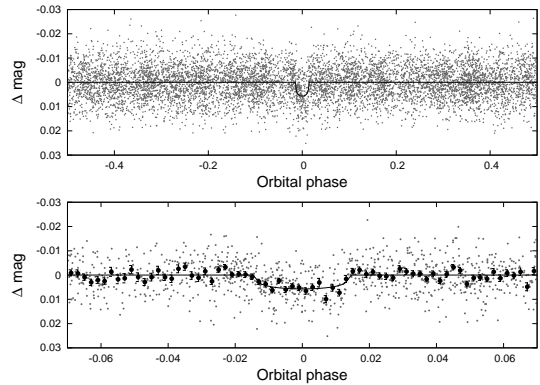}
\caption[]{
        Unbinned instrumental \band{r} \lc{} of \hatcur{} folded with
        the period $P = \hatcurLCPprec$\,days resulting from the global
        fit described in \refsecl{analysis}.  The solid line shows the
        best-fit transit model (see \refsecl{analysis}).  In the lower
        panel we zoom-in on the transit; the dark filled points here
        show the light curve binned in phase using a bin-size of
        0.002.\\
\label{fig:hatsouth}}
\end{figure}

The star \hatcur{} (2MASS \hatcurCCtwomass) was observed by the
HATSouth wide-field telescope network between UT 2011 March 24 and UT
2011 August 19.  These observations are summarized in
Table~\ref{tab:photobs}.  Observations were made from LCO
 in Chile, the H.E.S.S.~site in Namibia, and SSO in Australia. The data
were reduced to trend-filtered light curves following
\citet{penev:2013:hats1} and making use of the Trend Filtering
Algorithm \citep[TFA;][]{kovacs:2005:TFA}.  We searched the light
curves extracted from the images for periodic box-shaped transits using
the Box Least Squares \citep[BLS;][]{kovacs:2002:BLS} algorithm, and
identified a $P=\hatcurLCPshort{}$\,day signal with a depth of
5.1\,mmag in the light curve of \hatcur{} (Figure~\ref{fig:hatsouth};
the data are available in Table~\ref{tab:phfu}).

We searched the HATSouth light curve of \hatcur{} for other periodic
transit signals which would indicate the presence of additional
transiting planets in the system, but found no evidence for any such
signals.  We also searched for continuous periodic variability, due for
example to the presence of star-spots on \hatcur{}.  No significant
periodic signals are present in the TFA-processed light curve, and we
place an upper limit of 1\,mmag on the peak-to-peak amplitude of any
such variability.  We also checked for continuous periodic variability
in the light curve before applying the TFA filtering, and in this case
find that the maximum signal present ($P=39$\,days) has a peak-to-peak
amplitude of 2.7\,mmag, and is consistent with the level of
low-frequency systematic noise present in the pre-TFA light curves of
other bright stars in the field of \hatcur{}.

\ifthenelse{\boolean{emulateapj}}{
    \begin{deluxetable*}{llrrrr}
}{
    \begin{deluxetable}{llrrrr}
}
\tablewidth{0pc}
\tablecaption{
    Summary of Photometric Observations of \hatcur{}.
    \label{tab:photobs}
}
\tablehead{
    \colhead{Facility/Field \tablenotemark{a}} & 
    \colhead{Date Range} & 
    \colhead{Number of Points} & 
    \colhead{Median Cadence} & 
    \colhead{Filter} & 
        \colhead{Precision \tablenotemark{b}}\\
    \colhead{} & 
    \colhead{} & 
    \colhead{} &
    \colhead{(seconds)} &
    \colhead{} &
        \colhead{mmag}
}
\startdata
~~~~HS-2/G568 & 2011 Mar--2011 Aug & 5018 & 290 & Sloan~$r$ & 7.1 \\
~~~~HS-4/G568 & 2011 Jul--2011 Aug & 839 & 301 & Sloan~$r$ & 6.9 \\
~~~~HS-6/G568 & 2011 Mar--2011 May & 1627 & 295 & Sloan~$r$ & 6.3 \\
~~~~LCOGT1m+SBIG & 2014 Jun 12 & 110 & 136 & Sloan~$i^{\prime}$ & 2.0 \\
~~~~LCOGT1m+sinistro & 2014 Jun 16 & 49 & 167 & Sloan~$i^{\prime}$ & 1.3 \\
~~~~GROND & 2014 Jul 20 & 83 & 140 & Sloan~$g$ & 1.2 \\
~~~~GROND & 2014 Jul 20 & 83 & 140 & Sloan~$r$ & 0.9 \\
~~~~GROND & 2014 Jul 20 & 80 & 140 & Sloan~$i$ & 1.0 \\
~~~~GROND & 2014 Jul 20 & 80 & 140 & Sloan~$z$ & 1.0
\enddata
\tablenotetext{a}{For the HATSouth observations we list the HATSouth
  (HS)
  instrument used to perform the observations and the pointing on the
  sky. HS-2 is located at Las Campanas Observatory in Chile, HS-4 at
  the H.E.S.S.~gamma-ray telescope site in Namibia, and HS-6 at Siding
  Spring Observatory in Australia. Field G568 is one of 838 discrete
  pointings used to tile the sky for the HATNet and HATSouth
  projects. This particular field is centered at R.A.~14\,hr and
  Dec.~$-22.5^{\circ}$.}
\tablenotetext{b}{The r.m.s.~scatter of the residuals from our best
  fit transit model for each light curve at the cadence indicated in
  the Table.}
\ifthenelse{\boolean{emulateapj}}{
    \end{deluxetable*}
}{
    \end{deluxetable}
}

\subsubsection{Photometric follow-up}
\label{sec:phfu}

\begin{figure}[!ht]
\plotone{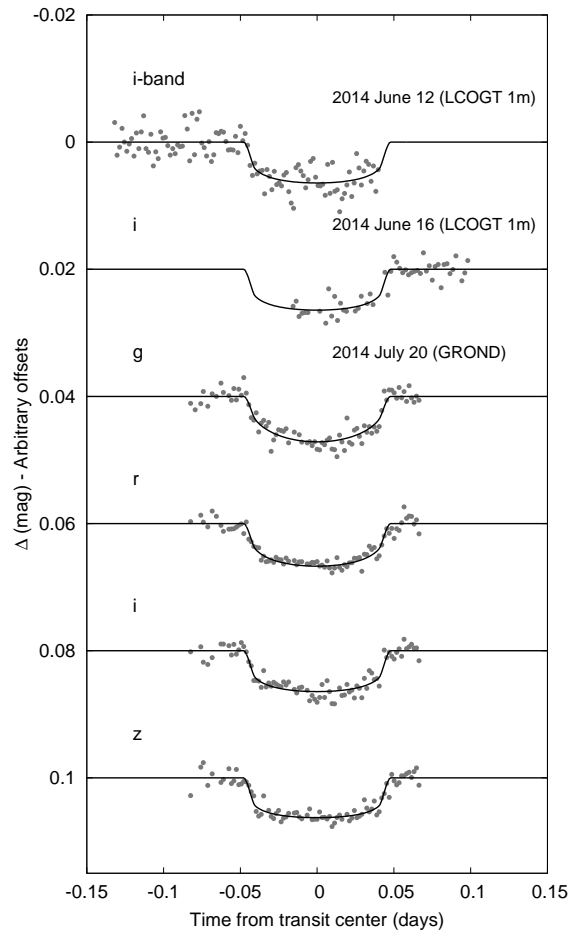}
\caption{
        Left: Unbinned follow-up transit \lcs{} of \hatcur{}.  The
        dates, filters and instruments used for each event are
        indicated.  The light curves have been detrended using the EPD
        process.  Curves after the first are shifted for clarity.  Our
        best fit is shown by the solid lines.
\label{fig:lc}}
\end{figure}

Photometric follow-up observations of \hatcur{} were performed using
the 1-m telescopes in the Las Cumbres Observatory Global Telescope
network (LCOGT) and the GROND instrument on the MPG~2.2\,m telescope at
La Silla Observatory (LSO) in Chile.  These observations are summarized
in Table~\ref{tab:photobs}.  Table~\ref{tab:phfu} provides the light
curve data, while the light curves are compared to our best-fit model
in Figure~\ref{fig:lc}.

Two transits were observed using the LCOGT~1\,m network
\citep{brown:2013:lcogt}.  The first, on UT 2014 June 12, was observed
using a 4K$\times$4K SBIG STX-16803 camera, with $0\farcs23$ pixels, at
the South African Astronomical Observatory (SAAO) station.  The second
transit, on UT 2014 June 16, was observed using the Sinistro camera at
Cerro Tololo Inter-American Observatory (CTIO) station in Chile, which
utilizes a 4K$\times$4K Fairchild CCD-486 back-side illuminated
detector with $0\farcs39$ pixels.  For both transits we used a Sloan
$i^{\prime}$ filter.  Because no apparent stellar neighbor is visible
on archival images, or through our own observations, we defocused the
telescope to reduce systematic errors in the photometry.  Standard CCD
calibrations were performed and light curves were extracted using
standard aperture photometry routines.

A single transit was observed on UT 2014 July 20 using the GROND
multi-filter instrument, providing simultaneous observations in the
Sloan $g$, $r$, $i$, and $z$ band-passes \citep{greiner:2008}. The
data were reduced to light curves following \citet{penev:2013:hats1}
and \citet{mohlerfischer:2013:hats2}.

\ifthenelse{\boolean{emulateapj}}{
        \begin{deluxetable*}{lrrrrr} }{
        \begin{deluxetable}{lrrrrr} 
    }
        \tablewidth{0pc}
        \tablecaption{Differential photometry of
        \hatcur\label{tab:phfu}} \tablehead{ \colhead{BJD} &
        \colhead{Mag\tablenotemark{a}} &
        \colhead{\ensuremath{\sigma_{\rm Mag}}} &
        \colhead{Mag(orig)\tablenotemark{b}} & \colhead{Filter} &
        \colhead{Instrument} \\ \colhead{\hbox{~~~~(2\,400\,000$+$)~~~~}}
        & \colhead{} & \colhead{} & \colhead{} & \colhead{} &
        \colhead{} } \startdata $ 55784.52625 $ & $  -0.00947 $ & $   0.00553 $ & $ \cdots $ & $ r$ &         HS\\
$ 55685.78275 $ & $   0.00387 $ & $   0.00490 $ & $ \cdots $ & $ r$ &         HS\\
$ 55682.59768 $ & $   0.00210 $ & $   0.00476 $ & $ \cdots $ & $ r$ &         HS\\
$ 55666.67146 $ & $   0.00587 $ & $   0.00552 $ & $ \cdots $ & $ r$ &         HS\\
$ 55762.23228 $ & $  -0.00517 $ & $   0.00421 $ & $ \cdots $ & $ r$ &         HS\\
$ 55784.52962 $ & $   0.01251 $ & $   0.00600 $ & $ \cdots $ & $ r$ &         HS\\
$ 55650.74655 $ & $   0.01291 $ & $   0.00467 $ & $ \cdots $ & $ r$ &         HS\\
$ 55749.49145 $ & $   0.00899 $ & $   0.00508 $ & $ \cdots $ & $ r$ &         HS\\
$ 55685.78731 $ & $   0.00285 $ & $   0.00494 $ & $ \cdots $ & $ r$ &         HS\\
$ 55682.60232 $ & $  -0.00122 $ & $   0.00467 $ & $ \cdots $ & $ r$ &         HS\\

        [-1.5ex]
\enddata \tablenotetext{a}{
     The out-of-transit level has been subtracted. For the HATSouth
     light curve (rows with ``HS'' in the Instrument column), these
     magnitudes have been detrended using the EPD and TFA procedures
     prior to fitting a transit model to the light curve.  Primarily as
     a result of this detrending, but also due to blending from
     neighbors, the apparent HATSouth transit depth is somewhat
     shallower than that of the true depth in the Sloan~$r$ filter (the
     apparent depth is 85\% that of the true depth).  For the follow-up
     light curves (rows with an Instrument other than ``HS'') these
     magnitudes have been detrended with the EPD procedure, carried out
     simultaneously with the transit fit (the transit shape is
     preserved in this process).
}
\tablenotetext{b}{
        Raw magnitude values without application of the EPD
        procedure.  This is only reported for the follow-up light
        curves.
}
\tablecomments{
        This table is available in a machine-readable form in the
        online journal.  A portion is shown here for guidance
        regarding its form and content. The data are also available on
        the HATSouth website at \url{http://www.hatsouth.org}.
} \ifthenelse{\boolean{emulateapj}}{ \end{deluxetable*} }{ \end{deluxetable} }

\subsection{Spectroscopy}
\label{sec:hispec}

Spectroscopic follow-up observations of \hatcur{} were carried out with
WiFeS on the ANU~2.3\,m telescope at SSO \citep{dopita:2007}, FEROS on
the MPG~2.2\,m \citep{kaufer:1998}, CORALIE on the Euler~1.2\,m at LSO
\citep{queloz:2001}, and HIRES on the Keck-I\,10\,m telescope at Mauna
Kea Observatory (MKO) in Hawaii \citep{vogt:1994}.

The WiFeS observations were carried out as part of our reconnaissance
to rule out false positives, and reduced and analyzed following
\citet{bayliss:2013:hats3} and \citet{zhou:2014:mebs}.  We obtained
three observations at a resolution of $R \equiv \lambda/\Delta\lambda =
7000$ to check for radial velocity variations in excess of $5$\,\kms,
and found that the observations were consistent with no variation at
this level.  A fourth spectrum at a resolution of $R = 3000$ was
obtained to provide an initial spectral classification and estimate of
the surface gravity.  Based on this observation we found that \hatcur{}
is a K dwarf star with an estimated effective temperature of $\teffstar
= 4700 \pm 300$\,K and a surface gravity of $\loggstar = 4.4 \pm 0.3$.

We obtained three $R = 48,000$ resolution FEROS spectra and five $R =
60,000$ resolution CORALIE spectra, which were reduced and analyzed
following \citet{jordan:2014:hats4}.  The RVs measured from these
spectra had a r.m.s.~scatter of 20\,\ms\ and were consistent with no
variation.  Based on these observations we concluded that \hatcurb{} is
a probable Neptune-mass planet, and continued observing it with the
larger aperture Keck-I telescope.  The stellar atmospheric parameters
estimated from the FEROS and CORALIE spectra are consistent with those
measured from WiFeS.  We also determined that the star is slowly
rotating (with $\vsini < 5$\,\kms), and, based on inspecting the
cross-correlation functions (CCFs), found no evidence for additional
stellar components in the spectra.

\begin{figure} [ht]
\plotone{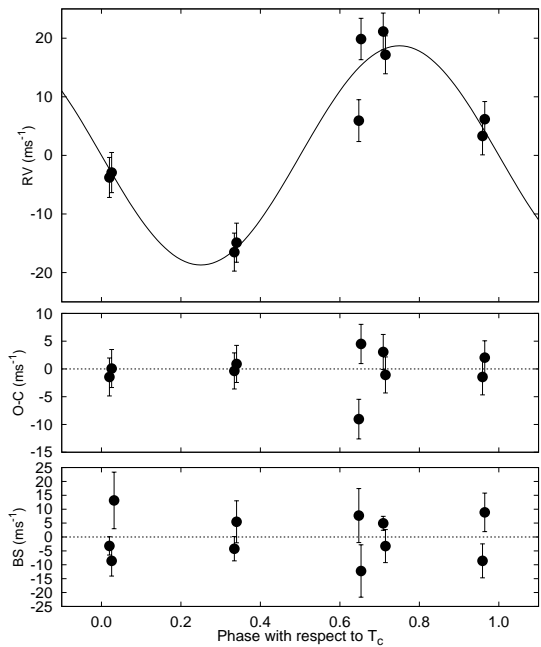}
\caption{
    {\em Top panel:} High-precision RV measurements for
    \hbox{\hatcur{}} from Keck/HIRES, together with our best-fit
    circular orbit model.  Zero phase corresponds to the time of
    mid-transit.  The center-of-mass velocity has been subtracted. 
    {\em Second panel:} Velocity $O\!-\!C$ residuals from the best-fit
    model.  The error bars for each instrument include the jitter which
    is determined in the fit.  {\em Third panel:} Bisector spans (BS),
    with the mean value subtracted.  Note the different vertical scales
    of the panels.\\
\label{fig:rvbis}}
\end{figure}

%
%
We observed \hatcur\ with the HIRES spectrometer (Vogt et al.~1994)
on the 10-m Keck-1 telescope using standard practices of the California
Planet Survey \citep[CPS; ][]{howard:2010}.
Our 10 observations through a
gas cell of molecular iodine (I$_2$) lasted 25 minutes each using the
``C2'' decker (14 $\times$ 0.86 arcsec slit).  In addition, we observed
\hatcur\ without the I$_2$ cell using the ``B3'' decker (14 $\times$ 0.57
arcsec slit) to record a template spectrum for the RV analysis and for
measuring high-precision stellar atmospheric parameters.  We computed
relative RVs using the \citet{butler:1996} method to model the I$_2$
$\times$ star spectrum.  RV errors were estimated from the uncertainty
on the mean of 700 spectral segments (each spanning $\sim$2 $\AA$) that
were separately analyzed for each observation.
We also measured spectral line bisector spans
(BSs) from these data following \citet{torres:2007:hat3}.  The final
RVs and BSs are provided in Table~\ref{tab:rvs} and are displayed in
Figure~\ref{fig:rvbis}.  The RVs vary in phase with the photometric
ephemeris and with a semiamplitude of $K = \hatcurRVK{}$\,\ms, while
the BSs are consistent with no variation and have an r.m.s.~scatter of
8.4\,\ms.

Based on the Keck/HIRES spectra, \hatcur\ is a quiet K dwarf, with
barely detectable Calcium HK activity ($\Savg = \hatcurSMES$, $\logrhk =
\hatcurSMElogrhk$; \reftab{stellar}).  Also, \hatcur\ has no
significant RV jitter (\reftab{planetparam}).  It is similar to other very
low activity K dwarfs, like HAT-P-26 \citep{hatp26}.

\ifthenelse{\boolean{emulateapj}}{
    \begin{deluxetable*}{lrrrrrr}
}{
    \begin{deluxetable}{lrrrrrr}
}
\tablewidth{0pc}
\tablecaption{
    Relative radial velocities and bisector span measurements of
    \hatcur{}.
    \label{tab:rvs}
}
\tablehead{
    \colhead{BJD} & 
    \colhead{RV\tablenotemark{a}} & 
    \colhead{\ensuremath{\sigma_{\rm RV}}\tablenotemark{b}} & 
    \colhead{BS} & 
    \colhead{\ensuremath{\sigma_{\rm BS}}} & 
        \colhead{Phase} &
        \colhead{Instrument}\\
    \colhead{\hbox{(2\,456\,000$+$)}} & 
    \colhead{(\ms)} & 
    \colhead{(\ms)} &
    \colhead{(\ms)} &
    \colhead{} &
        \colhead{} &
        \colhead{}
}
\startdata
$ 826.78970 $ & $    21.15 $ & $     2.07 $ & $    4.9 $ & $    2.5 $ & $   0.709 $ & Keck \\
$ 826.80787 $ & $    17.17 $ & $     2.21 $ & $   -3.3 $ & $    6.0 $ & $   0.715 $ & Keck \\
$ 827.78004 $ & $    -3.77 $ & $     2.46 $ & $   -3.2 $ & $    3.3 $ & $   0.020 $ & Keck \\
$ 827.79805 $ & $    -2.93 $ & $     2.47 $ & $   -8.6 $ & $    5.5 $ & $   0.026 $ & Keck \\
$ 827.81761 $\tablenotemark{c} & \nodata      & \nodata      & $   13.2 $ & $   10.2 $ & $   0.032 $ & Keck \\
$ 828.78151 $ & $   -16.51 $ & $     2.22 $ & $   -4.2 $ & $    4.4 $ & $   0.334 $ & Keck \\
$ 828.79940 $ & $   -14.90 $ & $     2.35 $ & $    5.5 $ & $    7.5 $ & $   0.340 $ & Keck \\
$ 829.77933 $ & $     5.93 $ & $     2.67 $ & $    7.7 $ & $    9.7 $ & $   0.648 $ & Keck \\
$ 829.79741 $ & $    19.86 $ & $     2.62 $ & $  -12.2 $ & $    9.4 $ & $   0.653 $ & Keck \\
$ 830.77141 $ & $     3.32 $ & $     2.19 $ & $   -8.6 $ & $    6.1 $ & $   0.959 $ & Keck \\
$ 830.78899 $ & $     6.17 $ & $     1.88 $ & $    8.9 $ & $    6.9 $ & $   0.965 $ & Keck \\

    [-1.5ex]
\enddata
\tablenotetext{a}{
        The zero-point of these velocities is arbitrary. An overall
        offset $\gamma_{\rm rel}$ is fitted to the Keck/HIRES
        velocities (\refsecl{analysis}), and has been subtracted.
}
\tablenotetext{b}{
        Internal errors excluding the component of
        astrophysical/instrumental jitter considered in
        \refsecl{analysis}.
}
\tablenotetext{c}{
        This HIRES observation was taken without the iodine cell to
        be used as a template. The RV is not measured for this
        observations, but the BS value is measured.
}
\ifthenelse{\boolean{emulateapj}}{
    \end{deluxetable*}
}{
    \end{deluxetable}
}

\section{Analysis}
\label{sec:analysis}

\begin{figure}[]
\plotone{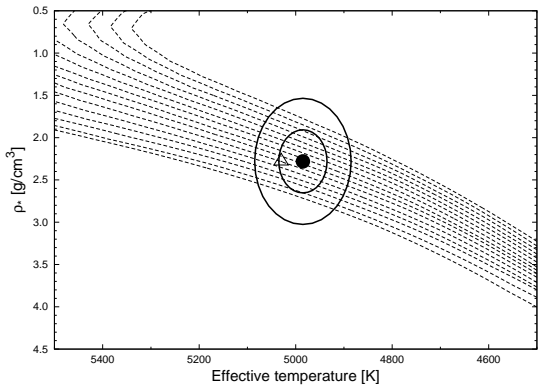}
\caption[]{
    Comparison between the measured values of \teffstar\ and \rhostar\
    (from SPC applied to the HIRES I$_{2}$-free template spectrum, and
    from our modeling of the light curves and RV data, respectively),
    and the Y$^{2}$ model isochrones from \citet{yi:2001}.  The
    best-fit values (dark filled circle), and approximate 1$\sigma$ and
    2$\sigma$ confidence ellipsoids are shown.  The values from our
    initial SPC iteration are shown with the open triangle.  The
    Y$^{2}$ isochrones are shown for ages of 0.2\,Gyr, and 1.0 to
    14.0\,Gyr in 1\,Gyr increments.
\label{fig:iso}}
\end{figure}

We analyzed the photometric and spectroscopic observations of \hatcur{}
to determine the parameters of the system using the standard procedures
developed for HATNet and HATSouth (see \citealp{bakos:2010:hat11}, with
modifications described by \citealp{hartman:2012:hat39hat41}).

High-precision stellar atmospheric parameters were measured from the
Keck/HIRES template spectrum using the Spectral Parameter
Classification program \citep[SPC;][]{buchhave:2012:spc}.  The
resulting \teffstar\ and \feh\ measurements were combined with the
stellar density \rhostar\ determined through our joint light curve and
RV curve analysis, to determine the stellar mass, radius, age,
luminosity, and other physical parameters, by comparison with the
Yonsei-Yale \citep[Y$^{2}$;][]{yi:2001} stellar evolution models (see
Figure~\ref{fig:iso}).  This provided a revised estimate of \loggstar\
which was fixed in a second iteration of SPC\@.  Our final adopted
stellar parameters are listed in Table~\ref{tab:stellar}.  We find that
the star \hatcur{} has a mass of \hatcurISOm\,\msun, a radius of
\hatcurISOr\,\rsun, and is at a reddening-corrected distance of
\hatcurXdist\,pc.  Note that the error on the distance does not take
into account the systematic errors due to uncertainties in the stellar
isochrones.  \hatcur\ is a main sequence star, and has a poorly
constrained age due to the limited stellar evolution expected within
the age of the universe at this stellar mass.

We also carried out a joint analysis of the HIRES RVs (fit using a
Keplerian orbit) and the HATSouth, LCOGT~1\,m, and GROND light curves (fit
using a \citealp{mandel:2002} transit model with fixed quadratic limb
darkening coefficients taken from \citealp{claret:2004}) to measure the
stellar density, as well as the orbital and planetary parameters.  This
analysis makes use of a differential evolution Markov Chain Monte Carlo
procedure \citep[DEMCMC;][]{terbraak:2006} to estimate the posterior
parameter distributions, which we use to determine the median parameter
values and their 1$\sigma$ uncertainties.  
We also varied the jitter as a free parameter in the fit following
\citet{hartman:2014:hat44to46}, where we use the empirical distribution
of the Keck/HIRES RV jitter values for low-activity cool main sequence
stars from \citet{wright:2005} to place a prior on the jitter.
The results are listed in
Table~\ref{tab:planetparam}.  We find that the planet \hatcurb{} has a
mass of \hatcurPPmlong\,\mjup, and a radius of \hatcurPPrlong\,\rjup. 
We also find that the observations are consistent with a circular
orbit.  When the eccentricity is allowed to vary in the fit, we find $e
= \hatcurRVecceneccen{}$ and a 95\% confidence upper-limit of
$e\hatcurRVeccentwosiglimeccen{}$.  The parameters listed in
Table~\ref{tab:planetparam} were determined assuming a fixed circular
orbit.

In order to rule out the possibility that \hatcur{} is a blended
stellar eclipsing binary system, we carried out a blend analysis of the
photometric data following \citet{hartman:2012:hat39hat41}.  We find
that a model consisting of single star with a transiting planet fits
the data better than any of the blended stellar eclipsing binary models
that we tested, but there are some blend models which cannot be
rejected with greater than 5$\sigma$ confidence based on the photometry
alone.  In these models the primary star in the background eclipsing
binary has a distance modulus that is no more than 2.1\,mag greater
than that of a brighter foreground star.  We simulated the
cross-correlation functions, RVs and BSs that would be measured with
HIRES/Keck-I for these blend scenarios and found that in almost all
cases the object would be easily identified as a composite stellar
system (the CCFs would be double-peaked, or the BSs and/or RVs would
vary by greater than 100\,\ms).  We searched the spectrum for the
spectral signature of a secondary star as described in
\citet{kolbl:2015}, and found no companions down to 1\% of the
brightness of the primary at separations in RV greater than
$\pm10\kms$.  There is a small region of parameter space where the BS
and RV scatter for the blended system would be comparable to the
measured scatters.  These are blends where the background binary has a
distance modulus that is between 1.95\,mag and 2.1\,mag greater than
that of the foreground star, and where the RV of the foreground star
differs by at least 50\,\kms\ from the systemic RV of the binary. 
Considering, however, that these blends can still be rejected with
4$\sigma$ confidence based on the photometry, and that the simulated
RVs do not show the sinusoidal variation in orbital phase that is
observed, we conclude that these blend scenarios do not provide a good
description of the observations, and that \hatcur{} is a transiting
planet system.

While we {\em can}\/ rule out the possibility that \hatcur{} is a
blended stellar eclipsing binary system, we cannot rule out the
possibility that \hatcur{} is a transiting planet system with a fainter
unresolved stellar companion.  We note that this is often the case for
many known transiting exoplanet systems.  We simulated this scenario as
well, and find that companions with masses up to that of the planet
host cannot be ruled out.  The highest spatial resolution images we
have available are from the HIRES guide camera based on which we can
rule out companions with $\Delta V < 4$ at a separation greater than
$1.5\arcsec$.  High mass companions would also need to have been
observed near conjunction for the system not to have been detected as a
spectroscopic binary.  Higher spatial resolution imaging and/or
additional RV monitoring would be needed to put tighter constraints on
any stellar companions.  If \hatcur{} does have a stellar companion,
the radius and mass of the planet \hatcurb\ would be somewhat larger
than what we infer here.

\ifthenelse{\boolean{emulateapj}}{
  \begin{deluxetable}{lcr}
}{
  \begin{deluxetable}{lcr}
}
\tablewidth{0pc}
\tabletypesize{\scriptsize}
\tablecaption{
    Stellar Parameters for \hatcur{} 
    \label{tab:stellar}
}
\tablehead{
    \multicolumn{1}{c}{~~~~~~~~Parameter~~~~~~~~} &
    \multicolumn{1}{c}{Value}                     &
    \multicolumn{1}{c}{Source}    
}
\startdata
\noalign{\vskip -3pt}
\sidehead{Identifying Information}
~~~~R.A.~(h:m:s)                      &  \hatcurCCra{} & 2MASS\\
~~~~Dec.~(d:m:s)                      &  \hatcurCCdec{} & 2MASS\\
~~~~R.A.p.m.~(mas/yr)                 &  \hatcurCCpmra{} & 2MASS\\
~~~~Dec.p.m.~(mas/yr)                 &  \hatcurCCpmdec{} & 2MASS\\
~~~~GSC ID                            &  \hatcurCCgsc{} & GSC                \\
~~~~2MASS ID                          &  \hatcurCCtwomass{} & 2MASS          \\
\sidehead{Spectroscopic properties}
~~~~$\teffstar$ (K)\dotfill         &  \hatcurSMEteff{} & SPC \tablenotemark{a}\\
~~~~Spectral type\dotfill           &  \hatcurISOspec & SPC                   \\
~~~~$\feh$\dotfill                  &  \hatcurSMEzfeh{} & SPC                 \\
~~~~$\vsini$ (\kms)\dotfill         &  \hatcurSMEvsin{} & SPC                 \\
~~~~$\gamma_{\rm RV}$ (\kms)\dotfill&  \hatcurRVgammaabs{} & FEROS            \\
~~~~$\Savg$   \dotfill              &  \hatcurSMES{}      & Keck/HIRES        \\
~~~~$\logrhk$ \dotfill              &  \hatcurSMElogrhk{} & Keck/HIRES        \\
\sidehead{Photometric properties}
~~~~$B$ (mag)\dotfill               &  \hatcurCCtassmB{} & APASS              \\
~~~~$V$ (mag)\dotfill               &  \hatcurCCtassmv{} & APASS              \\
~~~~$g$ (mag)\dotfill               &  \hatcurCCtassmg{} & APASS              \\
~~~~$r$ (mag)\dotfill               &  \hatcurCCtassmr{} & APASS              \\
~~~~$i$ (mag)\dotfill               &  \hatcurCCtassmi{} & APASS              \\
~~~~$J$ (mag)\dotfill               &  \hatcurCCtwomassJmag{} & 2MASS         \\
~~~~$H$ (mag)\dotfill               &  \hatcurCCtwomassHmag{} & 2MASS         \\
~~~~$K_s$ (mag)\dotfill             &  \hatcurCCtwomassKmag{} & 2MASS         \\
\sidehead{Derived properties}
~~~~$\mstar$ ($\msun$)\dotfill      &  \hatcurISOmlong{} & Y$^{2}$+\hatcurlumind{}+SPC \tablenotemark{b}\\
~~~~$\rstar$ ($\rsun$)\dotfill      &  \hatcurISOrlong{} & Y$^{2}$+\hatcurlumind{}+SPC         \\
~~~~$\rhostar$ (cgs)\dotfill        &  \hatcurISOrho{} & Y$^{2}$+\hatcurlumind{}+SPC         \\
~~~~$\loggstar$ (cgs)\dotfill       &  \hatcurISOlogg{} & Y$^{2}$+\hatcurlumind{}+SPC         \\
~~~~$\lstar$ ($\lsun$)\dotfill      &  \hatcurISOlum{} & Y$^{2}$+\hatcurlumind{}+SPC         \\
~~~~$M_V$ (mag)\dotfill             &  \hatcurISOmv{} & Y$^{2}$+\hatcurlumind{}+SPC         \\
~~~~$M_K$ (mag,\hatcurjhkfilset{})&  \hatcurISOMK{} & Y$^{2}$+\hatcurlumind{}+SPC         \\
~~~~Age (Gyr)\dotfill               &  \hatcurISOage{} & Y$^{2}$+\hatcurlumind{}+SPC         \\
~~~~$A_{V}$ (mag) \tablenotemark{c}\dotfill           &  \hatcurXAv{} & Y$^{2}$+\hatcurlumind{}+SPC\\
~~~~Distance (pc)\dotfill           &  \hatcurXdistred{} & Y$^{2}$+\hatcurlumind{}+SPC\\
\enddata
\tablenotetext{a}{
    SPC = ``Stellar Parameter Classification'' method based on
    cross-correlating high-resolution spectra against synthetic
    templates \citep{buchhave:2012:spc}.  These parameters rely
    primarily on SPC, but have a small dependence also on the iterative
    analysis incorporating the isochrone search and global modeling of
    the data, as described in the text.  }
\tablenotetext{b}{
    Isochrones+\hatcurlumind{}+SPC = Based on the Y$^{2}$ isochrones
    \citep{yi:2001},
    the stellar density used as a luminosity indicator, and the SPC
    results.
} 
\tablenotetext{c}{ Total \band{V} extinction to the star determined
  by comparing the catalog broad-band photometry listed in the table
  to the expected magnitudes from the
  Isochrones+\hatcurlumind{}+SPC model for the star. We use the
  \citet{cardelli:1989} extinction law.  }
\ifthenelse{\boolean{emulateapj}}{
  \end{deluxetable}
}{
  \end{deluxetable}
}

\ifthenelse{\boolean{emulateapj}}{
  \begin{deluxetable}{lr}
}{
  \begin{deluxetable}{lr}
}
\tabletypesize{\scriptsize}
\tablecaption{Parameters for the transiting planet \hatcurb{}.
\label{tab:planetparam}}
\tablehead{
    \multicolumn{1}{c}{~~~~~~~~Parameter~~~~~~~~} &
    \multicolumn{1}{r}{Value \tablenotemark{a}}                     
}
\startdata
\noalign{\vskip -3pt}
\sidehead{\Lc{} parameters}
~~~$P$ (days)             \dotfill    & $\hatcurLCP{}$              \\
~~~$T_c$ (${\rm BJD}$)    
      \tablenotemark{b}   \dotfill    & $\hatcurLCT{}$              \\
~~~$T_{14}$ (days)
      \tablenotemark{b}   \dotfill    & $\hatcurLCdur{}$            \\
~~~$T_{12} = T_{34}$ (days)
      \tablenotemark{b}   \dotfill    & $\hatcurLCingdur{}$         \\
~~~$\arstar$              \dotfill    & $\hatcurPPar{}$             \\
~~~$\zrstar$ \tablenotemark{c}              \dotfill    & $\hatcurLCzeta{}$\phn       \\
~~~$\rpl/\rstar$          \dotfill    & $\hatcurLCrprstar{}$        \\
~~~$b^2$                  \dotfill    & $\hatcurLCbsq{}$            \\
~~~$b \equiv a \cos i/\rstar$
                          \dotfill    & $\hatcurLCimp{}$           \\
~~~$i$ (deg)              \dotfill    & $\hatcurPPi{}$\phn         \\

\sidehead{Limb-darkening coefficients \tablenotemark{d}}
~~~$c_1,i$ (linear term)  \dotfill    & $\hatcurLBii{}$            \\
~~~$c_2,i$ (quadratic term) \dotfill  & $\hatcurLBiii{}$           \\
~~~$c_1,r$               \dotfill    & $\hatcurLBir{}$             \\
~~~$c_2,r$               \dotfill    & $\hatcurLBiir{}$            \\

\sidehead{RV parameters}
~~~$K$ (\ms)              \dotfill    & $\hatcurRVK{}$\phn\phn      \\
~~~$e$ \tablenotemark{e}  \dotfill    & $\hatcurRVeccentwosiglimeccen{}$ \\
~~~RV jitter (\ms) \tablenotemark{f}        \dotfill    & \hatcurRVjitter{}           \\

\sidehead{Planetary parameters}
~~~$\mpl$ ($\mjup$)       \dotfill    & $\hatcurPPmlong{}$          \\
~~~$\rpl$ ($\rjup$)       \dotfill    & $\hatcurPPrlong{}$          \\
~~~$C(\mpl,\rpl)$
    \tablenotemark{g}     \dotfill    & $\hatcurPPmrcorr{}$         \\
~~~$\rhopl$ (\gcmc)       \dotfill    & $\hatcurPPrho{}$            \\
~~~$\log g_p$ (cgs)       \dotfill    & $\hatcurPPlogg{}$           \\
~~~$a$ (AU)               \dotfill    & $\hatcurPParel{}$          \\
~~~$T_{\rm eq}$ (K) \tablenotemark{h}        \dotfill   & $\hatcurPPteff{}$           \\
~~~$\Theta$ \tablenotemark{i} \dotfill & $\hatcurPPtheta{}$         \\
~~~$\langle F \rangle$ ($10^{9}$\ergscmsq) \tablenotemark{i}
                          \dotfill    & $\hatcurPPfluxavg{}$       \\ [-1.5ex]
\enddata
\tablenotetext{a}{
    The adopted parameters assume a circular orbit. Based on the
    Bayesian evidence ratio we find that this model is strongly
    preferred over a model in which the eccentricity is allowed to
    vary in the fit. For each parameter we give the median value and
    68.3\% (1$\sigma$) confidence intervals from the posterior
    distribution.
}
\tablenotetext{b}{
    Reported times are in Barycentric Julian Date calculated directly
    from UTC, {\em without}\/ correction for leap seconds.
    \ensuremath{T_c}: Reference epoch of mid transit that
    minimizes the correlation with the orbital period.
    \ensuremath{T_{14}}: total transit duration, time
    between first to last contact;
    \ensuremath{T_{12}=T_{34}}: ingress/egress time, time between first
    and second, or third and fourth contact.
}
\tablenotetext{c}{
    Reciprocal of the half duration of the transit used as a jump
    parameter in our MCMC analysis in place of $\arstar$. It is
    related to $\arstar$ by the expression $\zrstar = \arstar
    (2\pi(1+e\sin \omega))/(P \sqrt{1 - b^{2}}\sqrt{1-e^{2}})$
    \citep{bakos:2010:hat11}.
}
\tablenotetext{d}{
    Values for a quadratic law, adopted from the tabulations by
    \cite{claret:2004} according to the spectroscopic (SPC) parameters
    listed in \reftabl{stellar}.
}
\tablenotetext{e}{
    The 95\% confidence upper-limit on the eccentricity from a model
    in which the eccentricity is allowed to vary in the fit.
}
\tablenotetext{f}{
    Error term, either astrophysical or instrumental in origin, added
    in quadrature to the formal RV errors for the listed
    instrument. This term is varied in the fit assuming a prior inversely 
    proportional to the jitter.
}
\tablenotetext{g}{
    Correlation coefficient between the planetary mass \mpl\ and
    radius \rpl\ determined from the parameter posterior distribution
    via $C(\mpl,\rpl) = \langle(\mpl - \langle\mpl\rangle)(\rpl -
    \langle\rpl\rangle)\rangle/(\sigma_{\mpl}\sigma_{\rpl})\rangle$, 
	where $\langle \cdot \rangle$ is the
    expectation value operator, and $\sigma_x$ is the standard
    deviation of parameter $x$.
}
\tablenotetext{h}{
    Planet equilibrium temperature averaged over the orbit, calculated
    assuming a Bond albedo of zero, and that flux is re-radiated from
    the full planet surface.
}
\tablenotetext{i}{
    The Safronov number is given by $\Theta = \frac{1}{2}(V_{\rm
    esc}/V_{\rm orb})^2 = (a/\rpl)(\mpl / \mstar )$
    \citep[see][]{hansen:2007}.
}
\tablenotetext{j}{
    Incoming flux per unit surface area, averaged over the orbit.
}
\ifthenelse{\boolean{emulateapj}}{
  \end{deluxetable}
}{
  \end{deluxetable}
}
%


\section{Discussion}
\label{sec:discussion}

\begin{figure*}[]
\plotone{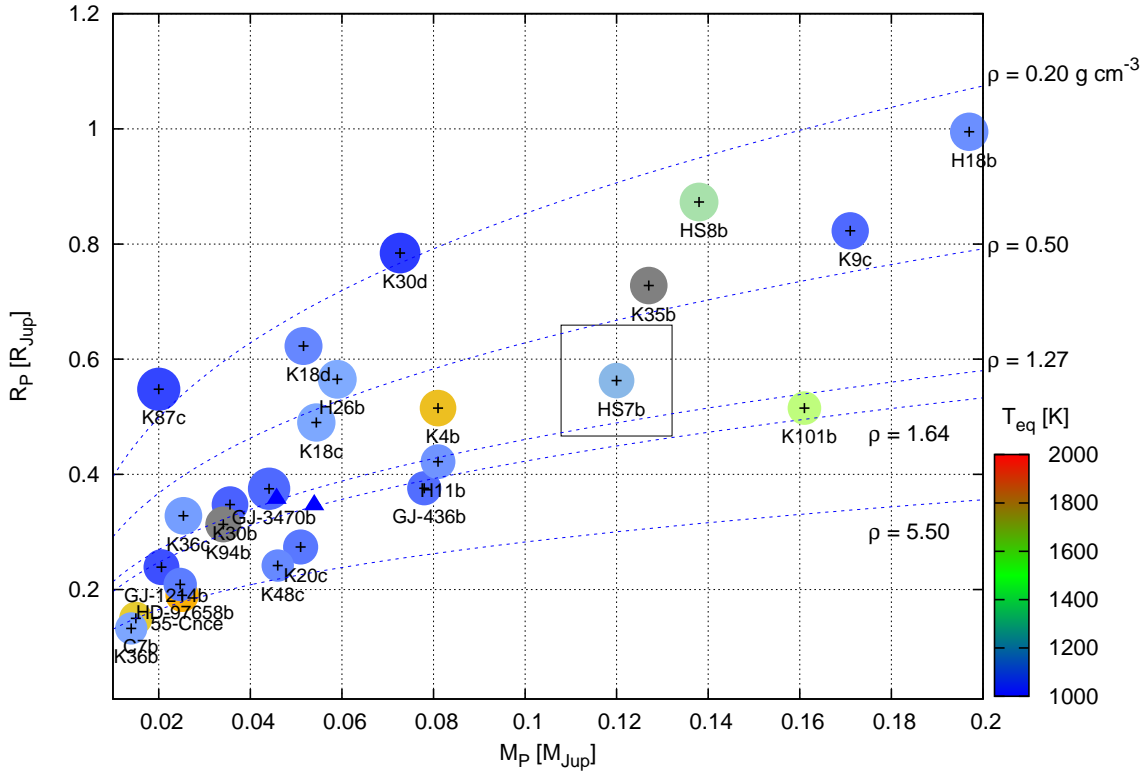}
\caption[]{
	Mass--radius diagram of super-Neptunes ($\mpl \le 0.18\mjup$) and
	super-Earths with accurately measured masses and radii ($<20$\%
	uncertainties).  Color indicates equilibrium temperature (with the
	palette of R,G,B = 2000, 1500, 1000\,K), while size is inversely
	proportional to surface gravity.  \hatcurb\ is also marked with a
	large box.  Isodensity lines for $\rho$ = 0.2, 0.5, 1.27 (Uranus),
	1.64 (Neptune), 5.5 (Earth) \gcmc\ are shown with dashed lines. 
	Neptune is marked with a blue triangle.  Abbreviations are: K:
	Kepler, H: HAT, HS: HATSouth, C: Corot.
\label{fig:mr}}
\end{figure*}

With its mass of $\mpl = \hatcurPPm\,\mjup$, HATS-7b is one of the very
few transiting and well characterized super-Neptunes known to date.  We
adopt a nomenclature, whereby planets with $0.18\,\mjup < \mpl \le
0.3\,\mjup$, i.e.~those more massive than halfway between Neptune and
Saturn, are referred to as sub-Saturns (esp.~those toward the lower end
of this mass range), whereas planets with $0.054\,\mjup < \mpl \le
0.18\,\mjup$ as super-Neptunes (esp.~those toward the higher mass end). 
This nomenclature is more or less consistent with previous literature,
e.g.~HAT-P-18b \citep{hartman:11:hatp18} is a called a sub-Saturn
($\mpl=0.197\,\mjup$), while Kepler-101b \citep{kep101b} is called a
super-Neptune ($\mpl=0.16\,\mjup$).  The mass vs.~radius of low mass
($\mpl < 0.2\,\mjup$) transiting planets with accurately determined
masses (here we use $<$20\% uncertainty) is shown in \reffig{mr}.  It
is noteworthy that only a handful of super-Neptunes are known (with
accurate parameters), and the regime between 0.1\,\mjup\ and
0.18\,\mjup\ is occupied by only 5 such objects.  These are Kepler-9c
\citep[K=12.34 mag; ][]{kep9c}, Kepler-101b \citep[K=12.0;
][]{kep101b}, the recently discovered HATS-8b \citep[K=12.6;
][]{bayliss:15:hats8}, Kepler-35b \citep[K=13.9; ][]{kep35b}, and
HATS-7b (K=10.97), the subject of this paper.  Among these planet
hosting stars, HATS-7 is by far the brightest, which will greatly
facilitate detailed studies.  All low-mass transiting planets shown in
\reffig{mr} were found by either {\em Kepler}\/ and {\em Corot}\/
(space-born surveys), HATNet and HATSouth (ground-based wide-field
surveys), the pointed ground-based survey MEarth \citep[GJ~1214b;
][]{mearth,gj1214b}, or they were found by the photometric follow-up of
planetary systems that were first discovered by RVs.

Well characterized planets and host stars in this regime are important
in exploring the correlations between various system parameters.  One
example is the host star metallicity -- planet occurrence rate
correlation, which has been known for a long time \citep{fischer:2005}
to exist for Jupiter size planets, and which correlation was found to
be weakened for small, $<4\rearth$ planets by \citet{buchhave:2012}. 
However, recent analysis by \citet{wang:2015} shows that the planet
occurrence rate--metallicity correlation is universal, including for
terrestrial planets.  Using an increased sample and new analysis,
\citet{buchhave:2014} find a correlation between host star metallicity
and planet radius, in the sense that the average metallicity of the
host star increases with planet size.  These authors also find three
populations of exoplanets (rocky, gas-dwarf, giant), as based on the
metallicity of the stars.  The discovery of HATS-7b around a metal rich
(\feh$=\hatcurSMEzfeh$) star, and the apparent lack of a hot Jupiter in
this system will help in refining such future analyses.

\begin{figure*}[]
\plotone{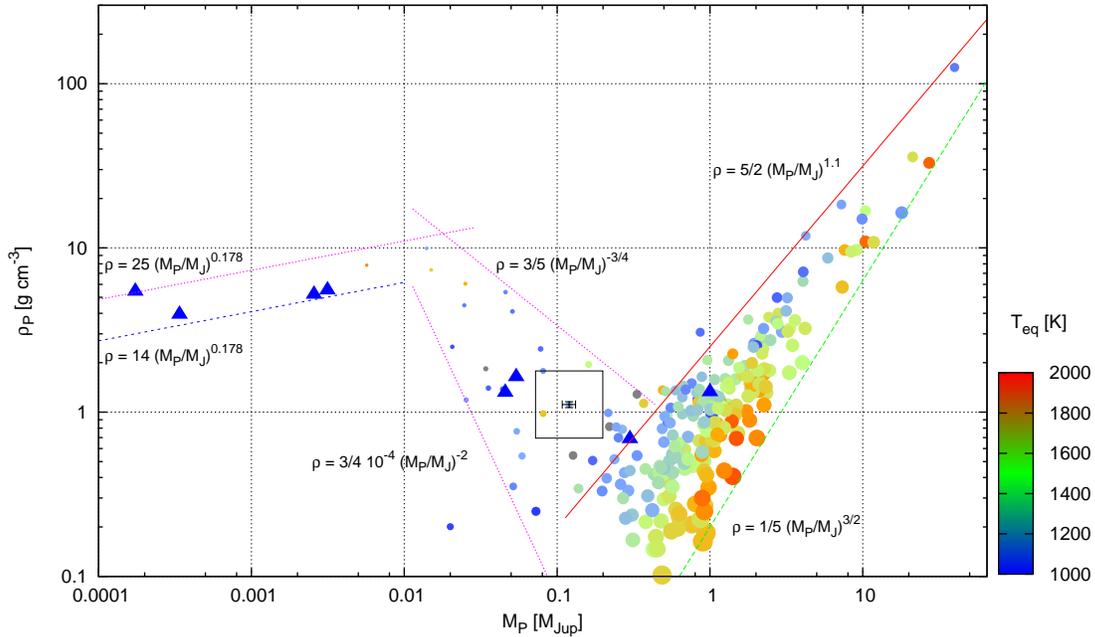}
\caption[]{
	Planetary mean density as a function of planetary mass for planets
	with mass measured to better than 20\% precision.  \hatcurb\ is
	also marked with a large box.  The size of the points scales with
	planetary radius, while the color indicates equilibrium temperature
	(with the palette of R,G,B = 2000, 1500, 1000\,K).  It is clearly
	visible that the less dense Jupiter-mass planets are predominantly
	`hot' (red).  The four thin lines indicate approximate boundaries
	of the current transiting extrasolar planet population in this
	space.  The significant outlier on the bottom left side is
	Kepler-87c \citep{ofir:2014:kep87c}, and the other significant
	outlier at $\mpl\approx 1\,\mjup$ and $\rho \approx 3\,\gcmc$ is
	WASP-59b \citep{hebrard:2013:wasp59}.  Solar system planets are
	marked with blue triangles.
\label{fig:rho}}
\end{figure*}

The mass regime of super Neptunes is important in studying the
transition from ice giants to gas giants, where, in the core accretion
scenario \citep[e.g.][]{mordasini:2015}, rapid accumulation of a
gaseous envelope is expected to start.  This transition is well
demonstrated in \reffig{rho}, which plots the bulk density of
transiting exoplanets as a function of their masses (again, only for
those with reliable parameters determined to better than 20\%
precision).  Mean density first {\em decreases} with increasing
planetary mass, as the planetary mass increases from super-Earths
($\sim0.01\,\mjup$) to Neptunes ($\sim0.1\,\mjup$).  Then, in the
super-Neptune/sub-Saturn regime, the trend reverses, as matter becomes
increasingly degenerate under high pressure, and density starts to
increase.

With the increasing number of well characterized transiting exoplanets,
we are gradually mapping out the parameter space these planets occupy. 
While it is customary to explore the relations between various
parameters (e.g.~mass, radius, equilibrium temperature), here we
examine the {\em boundaries}\/ spanned in the planetary mass ---
density domain.  The approximate envelopes of the distribution are:
\[
\mpl \lesssim 0.02\,\mjup:
\left\{
	\begin{array}{rl}
	\rhopl &\gtrsim 14\, (\mpl/\mjup)^{0.178}\\
	\rhopl &\lesssim 25\, (\mpl/\mjup)^{0.178}
	\end{array}
\right.\,,
\]
i.e.~for rocky planets without significant amount of volatiles, 
\[
\mpl \lesssim 0.4\,\mjup:
\left\{
	\begin{array}{rl}
	\rhopl &\gtrsim 3/4 \cdot 10^{-4}\, (\mpl/\mjup)^{-2}\\
	\rhopl &\lesssim 3/5\, (\mpl/\mjup)^{-3/4}
	\end{array}
\right.\,,
\]
i.e.~for Neptunes and Saturns, and
\[
\mpl \gtrsim 0.4\,\mjup:
\left\{
	\begin{array}{rl}
	\rhopl &\gtrsim 1/5\, (\mpl/\mjup)^{3/2}\\
	\rhopl &\lesssim 5/2\, (\mpl/\mjup)^{1.1}
	\end{array}
\right.
\]
for ice and gas giants ($\rhopl$ measured in \gcmc\ for all).  Here the
$\rhopl \propto (\mpl/\mjup)^{0.178}$ relation for the smallest planets
is taken from \citet{sotin:2007}, and is based on models.  The other
relations are approximate, and are not physically motivated, though it
is possible that physical explanations for them do exist.  Between
earths and super-earths, the mean density only increases slightly. 
Once $\mpl \gtrsim 0.02\,\mjup$ (6\,\mearth) is reached, the bulk
density decreases as a result of retaining more volatiles, and the
ability to form a more extended atmosphere.  Finally, at around $\mpl
\gtrsim 0.2\,\mjup$ ($\sim60\,\mearth$), the compression of matter gets
significant to result in an increasing mean density.

The lower limit for the density of gaseous planets (dashed line in
\reffig{rho}, $\propto \mpl^{3/2}$) is related to the inflation
mechanism of planets, which has not been fully solved.  It is
noteworthy that this relation holds over 2 orders of magnitude in mass,
from $\mpl\approx0.5\,\mjup$ to $50\,\mjup$.  The upper limit for the
density of rocky planets (dotted line, $\propto \mpl^{-3/4}$) is likely
to suffer from the bias against discovering compact rocky/iron planets,
due to their smaller transit signatures.  Also, while there are close
to 250 well characterized transiting exoplanets, considering the
dimensionality of the parameter space (e.g.~\mpl, \rpl, \mstar, $P$,
\feh, \teffstar, age), this is still a very small number to understand
their distribution.  Consequently, significant biases in the observed
distribution exist.  Probably the most important such bias is that the
majority of the planets plotted in \reffig{rho} have short periods. 
The confirmation and characterization of long period planets takes much
more time, and these systems are primarily discovered by {\em Kepler},
so the host stars are typically faint, further hindering follow-up
studies.

\begin{figure}[]
\plotone{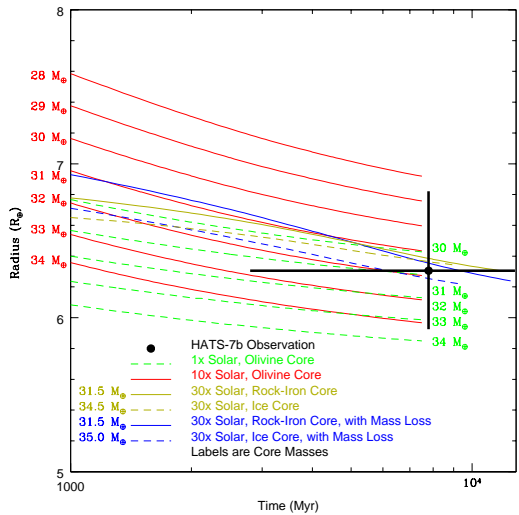}
\caption[]{
	Radius versus time for representative evolutionary models plotted
	against the observed radius and age estimates for HATS-7b.  Models
	are labeled with their corresponding core masses.  We include
	models with core masses from 30 to 34\,\mearth\ with a H$_2$-He
	atmosphere with solar metallicity, from 28 to 34 \mearth\ with an
	atmosphere with 10x solar metallicity, and four best-fit models
	with 30x solar metallicity.  In all cases, we fit the models to a
	(final) total mass of 38\,\mearth.
\label{fig:model}}
\end{figure}

The bulk composition of HATS-7b may be estimated by fitting theoretical
structural and evolutionary models to the observed mass and radius of
\hatcurPPme\,\mearth\ (\hatcurPPm\,\mjup) and \hatcurPPre\, \rearth\
(\hatcurPPr\,\rjup).  We compute a range of models for planets of
HATS-7b's mass for various possible parameters using the methodology of
\citet{howe:2015}.  If a two-layer partitioning with a rock-iron core
and a hydrogen-helium envelope is assumed, the best fit models suggest
a core mass of 31$\pm$4 \mearth\ and an envelope mass of 7$\pm$1.5
\mearth, that is, a hydrogen-helium fraction of 18$\pm$4\%.  This
result is consistent within uncertainty across a range of metallicity
and whether or not evaporative mass loss is incorporated in the model. 
More quantitatively, increasing the metallicity from $1\times$ solar to
$10\times$ solar increases the radius at 7.8 Gyr by about 0.2\,\rearth\
(0.018\,\rjup), approximately the same amount as decreasing the core
mass by 1\,\mearth.  This is also of the same order as the uncertainty
introduced by the uncertainty in age.  Models with significantly
smaller cores, $\lesssim$25\,\mearth, are ruled out because they under
no circumstances shrink to the observed radius of the planet over the
age of the system.  For comparison, final radii for the $10\times$
solar models with core masses of 0, 5, 10, 15, and 20 \mearth\ are
0.95, 0.87, 0.79, 0.71, and 0.65 \rjup, respectively (all larger than
the observed radius of \hatcurPPrshort\,\rjup), while a bare core of 38
\mearth\ has a radius of 0.202 \rjup\ if composed of rock-iron and
0.296 \rjup\ if composed of ice.  In the case of a pure ice core, the
best fit is a core mass of 34.5$\pm$4 \mearth\ and an envelope mass of
3.5$\pm$1.5 \mearth, that is, a hydrogen-helium fraction of 9$\pm$4\%,
similar to the composition of Uranus and Neptune \citep{helled:2011}. 
\reffig{model} shows radius versus time for representative evolutionary
models with core masses of 28-34 \mearth, encompassing the full range
of core masses consistent with observations.

\begin{figure*}[]
\plotone{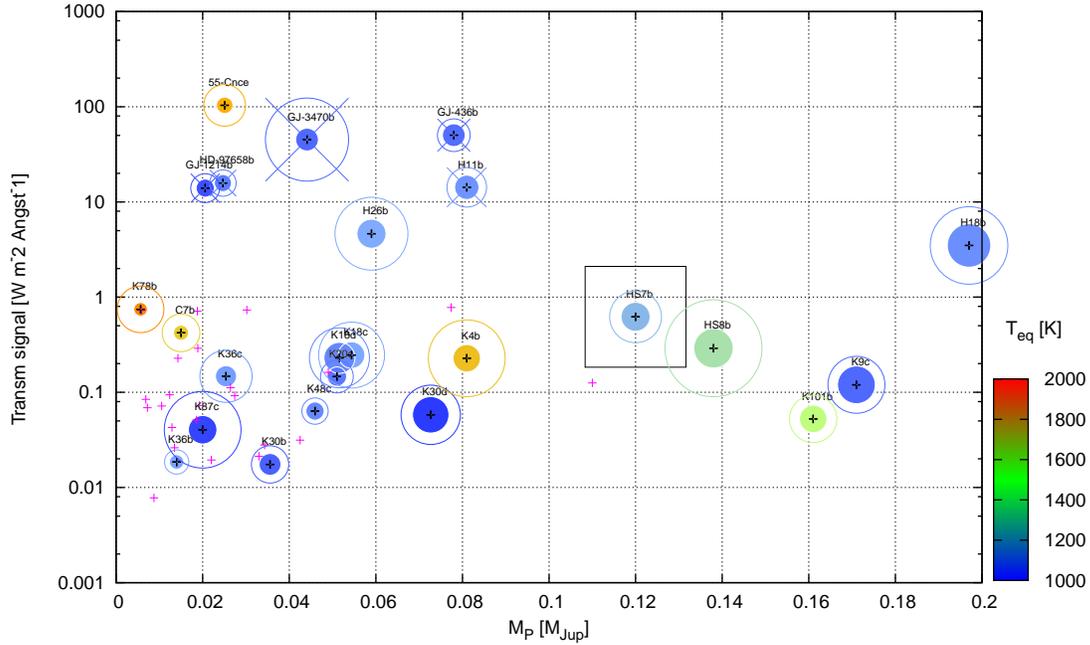}
\caption[]{
	Approximate detectability of a planet's atmosphere in transmission
	as a function of planetary mass for super-Neptunes ($\mpl \le
	0.18\mjup$) and super-Earths .  In calculating the scale height of
	the planet atmosphere we assume the molecular weight of pure
	molecular hydrogen, and the equilibrium temperature and surface
	gravity of the planet determined as from our analysis
	(\reftab{planetparam}).  The {\em fractional}\/ contribution of the
	transmission signal to the star light is approximated as $\delta =
	5 \times 2 \rpl H / \rstar^2$ \citep{perryman:2014}.  We then
	calculate the K-band flux of the star, and multiply with $\delta$,
	to come up with an approximate measure of the transmission signal. 
	We stress that this quantity does not take into account the
	detailed (expected) spectrum of the planetary atmosphere, but is an
	order of magnitude estimate of the signature.  The size of the
	filled circles scales with the radius of the planet (arbitrary
	scale), and the radius of the open circles scales with the scale
	height of the atmosphere.  Small plus symbols denote planets with
	uncertain mass or radius measurements (error $>$20\%). 
	Abbreviations are: K: Kepler, H: HAT, HS: HATSouth, C: Corot.
\label{fig:trans}}
\end{figure*}

While the mean density of exoplanets is an important constraint on
their bulk composition, detailed studies of the planetary atmospheres
can provide a direct measurement of the composition of the planet.  One
important way to detect such compositions is via transmission
spectroscopy.  In \reffig{trans} we show super-Neptunes to
super-Earths, including the sub-Saturn HAT-P-18b, with the expected
transmission signatures plotted against the planetary masses.  To
derive an approximate measure of the transmission signatures, we
approximated the $H$ scale height as $H = k_B T_{eq}/(\mu \gpl)$, where
$k_B$ is the Boltzmann constant, $T_{eq}$ is the equilibrium
temperature of the planets (assuming black bodies, and full
redistribution of heat), $\mu$ is the mean molecular weight, and \gpl\
is the planetary surface gravity.  To make the comparison simpler (and
because of lacking further information on their atmospheric
composition), we used $\mu=2$ (i.e.~$H_2$) for all planets, even though
the mean molecular weight of the atmospheres will be larger, and thus
our estimates are somewhat optimistic.  We then approximated the
fractional contribution of transmission through the planet's atmosphere
to the total stellar flux as $\delta = 5 \times 2 \rpl H / \rstar^2$
\citep{perryman:2014}, and multiplied this with the total estimated
K-band flux of the star.  Atmospheric transmission analysis has been
carried out for five of these planets, typically those in the top left
side of \reffig{trans} (those with low mass and high expected
transmission signal).  These are, in order of increasing planetary
mass: GJ-1214b \citep{kreidberg:14:gj1214}, \hd{97658b}
\citep{knutson:14:97658}, GJ~3470b \citep{ehrenreich:14:gj3470},
GJ~436b \citep{knutson:14:gj436} and HAT-P-11b
\citep{fraine:2014:hatp11}.  With the exception of HAT-P-11b, all
showed essentially featureless transmission spectra, indicating hazes,
clouds, or atmospheres with high molecular weight.  The only Neptune
mass planet with features in its transmission spectrum is HAT-P-11b,
where the signature of water was detected \citep{fraine:2014:hatp11}. 
Among the super Neptunes, HATS-7b is the most promising target for
detecting its atmosphere.  Also, such measurements will be facilitated
by the quiet host star; high resolution spectroscopy of HATS-7 shows
very low RV jitter, low \vsini, and no chromospheric activity. 
Altogether, it is scientifically compelling to pursue transmission
spectroscopy for the recently discovered super Neptunes HATS-7b and
HATS-8b, and the sub-Saturn HAT-P-18b, where relatively large signals
are expected.  Perhaps a trend from featureless spectra to feature-rich
transmission (observed for certain hot Jupiters) will be observed with
increasing planetary mass through the super-Neptune regime.

\acknowledgements 

\paragraph{Acknowledgements}
Development of the HATSouth project was funded by NSF MRI grant
NSF/AST-0723074, operations have been supported by NASA grants
NNX09AB29G and NNX12AH91H, and follow-up observations received partial
support from grant NSF/AST-1108686.
A.J.\ acknowledges support from FONDECYT project 1130857, BASAL CATA
PFB-06, and project IC120009 ``Millennium Institute of Astrophysics
(MAS)'' of the Millenium Science Initiative, Chilean Ministry of
Economy.  R.B.\ and N.E.\ are supported by CONICYT-PCHA/Doctorado
Nacional.  R.B.\ and N.E.\ acknowledge additional support from project
IC120009 ``Millenium Institute of Astrophysics (MAS)'' of the
Millennium Science Initiative, Chilean Ministry of Economy.  V.S.\
acknowledges support form BASAL CATA PFB-06.  
This work is based on observations made with ESO Telescopes at the La
Silla Observatory.
This paper also uses observations obtained with facilities of the Las
Cumbres Observatory Global Telescope.
Work at the Australian National University is supported by ARC Laureate
Fellowship Grant FL0992131.
We acknowledge the use of the AAVSO Photometric All-Sky Survey (APASS),
funded by the Robert Martin Ayers Sciences Fund, and the SIMBAD
database, operated at CDS, Strasbourg, France.
Operations at the MPG~2.2\,m Telescope are jointly performed by the
Max Planck Gesellschaft and the European Southern Observatory.  The
imaging system GROND has been built by the high-energy group of MPE in
collaboration with the LSW Tautenburg and ESO\@.  
The authors wish to
recognize and acknowledge the very significant cultural role and
reverence that the summit of Mauna Kea has always had within the
indigenous Hawaiian community. We are most fortunate to have the
opportunity to conduct observations from this mountain.
G.~\'A.~B.~wishes to thank the warm hospitality of Ad\`ele and Joachim Cranz
at the farm Isabis, supporting the operations and service missions of
HATSouth.

\clearpage
\bibliographystyle{apj}
\bibliography{hatsbib}

\end{document}